\documentclass[a4paper,fleqn,usenatbib]{mnras}

\usepackage{newtxtext,newtxmath}
\usepackage[T1]{fontenc}
\usepackage{ae,aecompl}

\usepackage{float}

\usepackage{graphicx}	% Including figure files
\usepackage{amsmath}	% Advanced maths commands
\usepackage{amssymb}	% Extra maths symbols
\usepackage{lineno}
\usepackage{afterpage}
\setcitestyle{notesep={}}
\title[SAMI: S0 Formation]{The SAMI Galaxy Survey: A Range in S0 Properties Indicating Multiple Formation Pathways}

\author[S. Deeley et al.]{
Simon Deeley$^{1}$\thanks{E-mail: s.deeley@uq.edu.au},
Michael J. Drinkwater$^{1}$, Sarah M. Sweet$^{1,2,3}$, Jonathan Diaz$^{4}$, 
\newauthor Kenji Bekki$^{4}$, Warrick J. Couch$^{2}$, Duncan A. Forbes$^{2}$, Joss Bland-Hawthorn$^{5}$,
\newauthor Julia J. Bryant$^{3,5,6}$,
 Scott Croom$^{3,5}$, Luca Cortese$^{3,4}$, Jon S. Lawrence$^{7}$, Nuria Lorente$^{8}$,
\newauthor Anne M. Medling$^{9,10}$\thanks{Hubble Fellow}, Matt Owers$^{11,12}$, Samuel N. Richards$^{13}$ 
\newauthor and Jesse van de Sande$^{3,5}$ 
\\
${}^1$School of Mathematics and Physics, University of Queensland, Brisbane, Queensland 4072, Australia\\
${}^2$Centre for Astrophysics \& Supercomputing, Swinburne University, Hawthorn, VIC 3122, Australia \\
${}^3$ARC Centre of Excellence for All Sky Astrophysics in 3 Dimensions (ASTRO 3D)\\
${}^4$International Centre for Radio Astronomy Research, The University of Western Australia, 35 Stirling Highway, Crawley, Western Australia, 6009, Australia \\
${}^5$Sydney Institute for Astronomy, School of Physics, The University of Sydney, NSW 2006, Australia \\
${}^6$Australian Astronomical Optics, AAO-USydney, School of Physics, University of Sydney, NSW 2006, Australia\\ 
${}^7$Australian Astronomical Optics - Macquarie, Macquarie University, NSW 2109, Australia\\
${}^8$AAO-MQ, Faculty of Science \& Engineering, Macquarie University. 105 Delhi Rd, North Ryde, NSW 2113, Australia\\
${}^{9}$Research School of Astronomy and Astrophysics, Australian National University, Canberra, ACT 2611, Australia\\
${}^{10}$Ritter Astrophysical Research Center University of Toledo Toledo, OH 43606, USA\\
${}^{11}$Department of Physics and Astronomy, Macquarie University, NSW 2109, Australia\\
${}^{12}$Astronomy, Astrophysics and Astrophotonics Research Centre, Macquarie University, Sydney, NSW 2109, Australia\\
${}^{13}$SOFIA Science Center, USRA, NASA Ames Research Center, Building N232, M/S 232-12, P.O. Box 1, Moffett Field, CA 94035-0001, USA\\
}

\date{Accepted XXX. Received YYY; in original form ZZZ}

\pubyear{2019}

\begin{document}
%\linenumbers
\label{firstpage}
\pagerange{\pageref{firstpage}--\pageref{lastpage}}
\maketitle
%\raggedbottom
\setlength{\parskip}{0pt}
% Abstract of the paper  NOTE 250 word limit ****
\begin{abstract}
 
It has been proposed that S0 galaxies are either fading spirals or the result of galaxy mergers. The relative contribution of each pathway, and the environments in which they occur remains unknown. Here we investigate stellar and gas kinematics of 219 S0s in the SAMI Survey to look for signs of multiple formation pathways occurring across the full range of environments. We identify a large range of rotational support in their stellar kinematics, which correspond to ranges in their physical structure. We  find that pressure-supported S0s with $v/\sigma$ below 0.5 tend to be more compact and feature misaligned stellar and gas components, suggesting an external origin for their gas. We postulate that these S0s are consistent with being formed through a merger process. Meanwhile, comparisons of ellipticity, stellar mass and S\'ersic index distributions with spiral galaxies shows that the rotationally supported S0s with $v/\sigma$ above 0.5 are more consistent with a faded spiral origin. In addition, a simulated merger pathway involving a compact elliptical and gas-rich satellite results in an S0 that lies within the pressure-supported group. We conclude that two S0 formation pathways are active, with mergers dominating in isolated galaxies and small groups, and the faded spiral pathway being most prominent in large groups ($10^{13}<\rm{M_{halo}}<10^{14}$).
\end{abstract}


%\begin{keywords}

%\end{keywords}

%%%%%%%%%%%%%%%%%%%%%%%%%%%%%%%%%%%%%%%%%%%%%%%%%%

%%%%%%%%%%%%%%%%% BODY OF PAPER %%%%%%%%%%%%%%%%%%
\section{Introduction}
\label{introduction}

The visual morphological class of S0 galaxies was originally defined by \citet{1926ApJ....64..321H} as an intermediate class between early-type elliptical galaxies and late-type spirals. S0s feature a central bulge surrounded by a disk, reminiscent of most spiral galaxies. However, the gas content of S0s is significantly lower, they exhibit no (or extremely weak) spiral structure, and their redder colours are similar to that of ellipticals \citep{1978ApJ...225..742S,1970SoPh...15..288F}. While some S0s have been observed to have ongoing star formation, the rate of star formation is typically insignificant compared to that seen in spirals \citep{2009ApJ...695....1T}. 

The relative fraction of S0 galaxies is observed to increase with density in nearby rich clusters \citep{1980ApJ...236..351D}, with S0s becoming the most common galaxy in the densest clusters. Observations of rich clusters at redshifts up to $z\sim0.5$ indicate that there has been a buildup in the S0 population in such systems over the last 5 Gyr, with a corresponding decrease in their spiral population (e.g. \citet{1997ApJ...490..577D}). Significant numbers of S0 galaxies are also seen within smaller galaxy groups, where the evolution of the S0 fraction with redshift is more rapid compared to more massive groups and clusters (\citet{2010ApJ...711..192J}, \citet{2009ApJ...697L.137P}). The location of S0 formation is often assumed to be within the large clusters they are found in, however observations by \citet{2010ApJ...711..192J} are consistent with a scenario whereby S0s form within smaller groups which later fall in and merge into the larger cluster systems. The environments of small groups may be the most conductive to S0 formation, where galaxy-galaxy interactions are expected to be more important than galaxy-environment interactions \citep{1992ARA&A..30..705B}.
 
The properties of S0 galaxies in lower density environments also differ to those in clusters. S0 galaxies in high density environments such as the Virgo cluster appear to have cores bluer (and therefore younger) then the surrounding disk \citep{2014MNRAS.441..333J}, while S0s in the low density field environment feature redder cores relative to their disks \citep{2017MNRAS.466.2024T}. This indicates either a different formation pathway for S0s in the field, or S0 evolution after initial formation outside the cluster. 

In addition, studies of S0 galaxies in field and cluster environments have revealed that those with significant gaseous emission are mainly found in the field and cluster outskirt regions. This suggests that either the gas is stripped as the galaxy falls into a cluster, or that the elevated gas content arises from mergers or inflow of fresh gas which may be more common in the lower density environments \citep{2014MNRAS.440.3491J}. 
Decoupled gas kinematics in isolated S0 galaxies also supports an external origin for the gas content in low density environments \citep{2014MNRAS.438.2798K}. Surveys of emission-selected S0s have revealed that a number of these objects feature a ring of gas emission accompanied by star formation activity \citep{2014MNRAS.439..334I}. 

Due to the common bulge/disk structure of spiral and S0 galaxies, and the anti-correlation seen between the fraction of the two classes as a function of local density, it is commonly assumed that S0s originate from spiral galaxies. Under that assumption, various physical processes have been proposed to explain the properties and locations of S0 galaxies. Formation via ram pressure stripping \citep{1972ApJ...176....1G} occurs when a spiral galaxy travels through the intergalactic medium of a cluster \citep{2012ApJ...750L..23O}. The interaction between the gas and dust in the spiral galaxy with that of the medium strips gas out of the galaxy (see for example \citet{2000Sci...288.1617Q}). Direct evidence supporting ram pressure stripping includes jellyfish galaxies, where new stars are seen forming in the streams of gas stripped out of a spiral galaxy e.g. \citet{2014ApJ...781L..40E}. Gravitational interactions in clusters \citep{1980ApJ...237..692L} and groups \citep{2011MNRAS.415.1783B} between a gas rich spiral and a neighbouring galaxy can thicken the disk and disrupt its spiral structure. Additional spiral - S0 processes include spiral-spiral mergers \citep{2017A&A...604A.105T} and halo stripping \citep{2002ApJ...577..651B}. 

A new pathway in lower density environments was recently proposed by \citet{2018MNRAS.tmp..724D}, where an isolated, high-redshift compact elliptical (cE) experiences a merger with a smaller gas-rich disk galaxy. The modelling showed that the gas from the satellite forms a ringed disk around the core of the cE and a burst of star formation is triggered within it. While present, the gas ring features more coherent rotation relative to the stars, and therefore has higher $v/\sigma$ values (where v is the rotational velocity and $\sigma$ is the velocity dispersion). Once the gas has been depleted and star formation has ceased, the end result is an older core, a younger disk and an extended halo, giving a negative age gradient with increasing radius. The remnants of the original elliptical leave a kinematic imprint on the resulting S0 galaxy, in the form of a flattened velocity profile. The stellar $v/\sigma$ in the resulting S0 is found to be near zero in the centre and remain below 0.5 out to ${2 \rm R_{e}}$, which is in contrast to spiral merger scenarios \citep[$\sim1$,][]{1998ApJ...502L.133B} and fading spiral models \citep[stellar $v/\sigma$ $\sim2-10$,][]{2000Sci...288.1617Q}. Stellar $v/\sigma$ profiles can therefore be used to identify S0s which may potentially originate from the cE + disk merger pathway. In addition, the negative population gradient expected from the cE + disk pathway contrasts with the positive age gradient expected for spiral fading pathways, providing an additional distinguishing feature between the pathways. 

Recent observations have suggested there may be separate populations of S0 galaxies with different formation pathways. In a study of 21 S0s in extreme field and cluster environments, \citet{2020MNRAS.492.2955C} found that those in clusters are more rotationally supported than those in the field, arguing that cluster S0s are formed via rapid removal of gas (e.g. ram pressure stripping), whereas mergers are involved in the formation of field S0s. In contrast, \citet{2018MNRAS.476.2137R} studied 9 S0s in low density environments and found that seven were consistent with a faded-spiral origin and two with a merger origin. \citet{2019ApJ...880..149P} compared nine S0s in the CALIFA survey to passive spirals and found a larger range of stellar population properties in the S0s relative to the passive spirals, indicating again that faded spirals may only be one of multiple S0 formation pathways. Recent work has therefore suggested multiple pathways may be active, however the conclusions have been limited by small sample sizes and restricted environmental coverage. 

To more clearly separate these potential S0 formation pathways, resolved kinematic data is needed in order to investigate the kinematic properties as a function of radius. New integral field surveys such as CALIFA \citep{2012A&A...538A...8S} and the SAMI Galaxy Survey \citep{2018MNRAS.475..716G} provide 2D spectral data for many galaxies. The advantage of SAMI is the coverage over a wide range of environments, allowing us to investigate the 2D kinematics of many S0s across the different environments in which they occur.  

In this paper, the aims are firstly to corroborate the above claims that S0 galaxies have formed through both the faded spiral and merger pathways, and secondly to test the cE + disk scenario by comparing the properties of the resulting S0 galaxy in simulations with those of observed S0s. In Section 2 we describe the galaxy sample used in this work. Section 3 outlines the methods used to analyse the observational data. Section 4 shows the derived kinematic and gas emission properties of observed S0 galaxies and compares them to the simulations. In Section 5 we discuss the results and in Section 6 we present our summary and conclusions. 

We assume a ${\rm \Lambda CDM}$ cosmology with ${\rm \Omega_{m}}=0.3$, ${\rm \Omega_{\lambda}}=0.7$ and ${\rm H_{0}=70 km s^{-1} kpc^{-1}}$.

\section{Data}
\label{data}

\subsection{The SAMI survey}

The observational data products used in this work were created by the SAMI Galaxy Survey \citep{2015MNRAS.447.2857B}. The Sydney-AAO Multi-object Integral field spectrograph \citep[SAMI;][]{2012MNRAS.421..872C} is mounted at the prime focus of the 3.9m the Anglo-Australian Telescope, which provides a 1 degree diameter field of view. SAMI uses 13 fused fibre bundles \citep[Hexabundles;][]{2011OExpr..19.2649B, 2014MNRAS.438..869B} with a high (75 percent) fill factor. Each bundle contains 61 fibres of 1.6 arcsec diameter resulting in each hexabundle having a diameter of 15 arcsec. The hexabundles, as well as 26 sky fibres, are plugged into pre-drilled plates using magnetic connectors. SAMI fibres are fed to the double-beam AAOmega spectrograph \citep{2006SPIE.6269E..0GS}. For the SAMI Galaxy Survey its 570V grating was used with the blue arm (3700-5700A), giving a resolution of R=1730 (${\rm \sigma=74 km/s}$), and the R1000 grating with the red arm (6250-7350A) giving a resolution of R=4500 (${\rm \sigma=29 km/s}$) \citep{2017ApJ...835..104V}. At least six pointings on each galaxy are weighted and combined to produce data cubes with a pixel scale of $0.5\times0.5$ arcseconds \citep{2015MNRAS.446.1567A, 2015MNRAS.446.1551S}. Here we use data products released in Data Release 2 \citep{2018MNRAS.481.2299S}.

\subsection{Galaxy sample}

The SAMI survey was designed to cover a wide range of environments to allow for studies of galaxy properties as a function of local environment. To achieve this, the survey targeted two regions. The first coincides with the Galaxy and Mass Assembly (GAMA) survey \citep{2011MNRAS.413..971D}, which covers 286 square degrees and has a limiting $r-$band magnitude of 19.8. Since GAMA includes limited coverage of high density cluster regions, SAMI targeted additional galaxies within eight large clusters in order to achieve a more complete coverage of environments \citep{2017MNRAS.468.1824O}. In both regions, galaxies were selected using the same volume-limited mass limits. Table~\ref{data_table} displays the number of S0s in each region of the SAMI survey (see below for details on classification). 

\subsection{Environment}
\label{environment_methof}

The high completeness of the GAMA survey enabled a complete group catalogue to be constructed for galaxies in this region \citep{2011MNRAS.416.2640R}. We used this catalogue to identify the host environment of each SAMI galaxy, depending on the mass of their host group halo. We define host group masses in the range $10^{11} < {\rm M}_{\odot} < 10^{13}$ as low-mass groups and those with ${\rm M}_{\odot} > 10^{13}$ as high-mass groups. Galaxies not associated with a group are identified as field galaxies, and those located in the eight cluster regions outside of the GAMA field are identified as cluster galaxies.

High resolution images of all SAMI galaxies within the GAMA region were taken using the Hyper-Suprime Cam on the Subaru Telescope in Hawaii \citep{2018PASJ...70S...8A}. These images were taken in five colour bands - $g,i,r,Y$ and $z$. The depth of the imaging extends to $i$-band magnitudes of 28, and the images have a resolution of 0.168 arcseconds per pixel. The high depth and resolution of these images allows for the identification of both prominent and very faint structural features such as disks, shells and tidal arms within and around the galaxy, providing clues to their evolutionary history.  

For galaxies in the GAMA region, the S\'ersic index, ellipticity and effective radius are derived from a single component fit of SDSS imagery using the code GALFIT3 \citep{2010AJ....139.2097P, 2011MNRAS.413..971D}. In the cluster regions, these physical parameters are derived from r-band images from the SDSS and VLT Survey telescope/ATLAS surveys using the code PROFIT \citep{2017MNRAS.466.1513R, 2019ApJ...873...52O}. The stellar masses in both regions were derived from mass/light ratios based on $g-i$ colours \citep{2015MNRAS.447.2857B}.

\subsection{Morphological Classification}
\label{galaxy sample}

Galaxies in the SAMI Survey were visually classified by a subset of survey team members people using Sloan Digital Sky Survey imaging \citep{2016MNRAS.463..170C}. Firstly, the galaxy was classed as an early or late type galaxy based on the absence or presence of a clear disk, spiral arm features and areas of star formation. In the second step, galaxies in the early-type category were classed as lenticular (S0) if a disk component was visible or elliptical (E) otherwise. Late-type galaxies were classed as either early spirals (Sa, bulge present) or late type spirals (Sp, no bulge visible). Classes were assigned only if 66 percent of the classifiers agreed on the type. Intermediate classes (e.g. E/S0) were assigned to cases where 66 percent agreed it was one or the other. For all other cases, the galaxy was classed as unknown. In our analysis, the Sa, Sa/Sp and Sp classes are grouped into a single category labelled spirals and E/S0, S0 and S0/Sa are grouped together as S0s.

\section{Methods}
\label{methods}

In this section we describe how we derive the physical parameters and radial kinematic profiles used in this work, as well as the method used to derive the stellar age profiles. 

\subsection{Kinematics}

In order to allow for direct comparisons of S0 kinematics with the Diaz et al. (2018) simulation, we use the SAMI kinematic maps to derive radial profiles of the inclination-corrected velocity, the velocity dispersion and their ratio for the stellar and gas components.

The method used to derive stellar kinematic maps for each galaxy is detailed in \citet{2017ApJ...835..104V}. Briefly, both the red and blue arms are used for fitting the stellar kinematics, with the red arm convolved to match the spectral resolution in the blue arm. Elliptical annuli are used to derive the optimal stellar template for each region of the galaxy in order to mitigate strong radial gradients. The penalised pixel fitting code \citep[pPXF][]{2004PASP..116..138C} is then used on each spaxel, producing two-moment and four-moment data products. Here we use the two-moment velocity and velocity dispersion, derived from Gaussian line of sight velocity dispersion fits to each spaxel. The fitting tool LZIFU \citep{2016Ap&SS.361..280H} is used to derive the emission line physics for each spaxel \citep[see][ for full details]{2018MNRAS.475..716G}. LZIFU simultaneously fits all the strong emission lines with both a single-component Gaussian fit and a multi-component fit; here we use the single component fit to investigate the emission line and gas kinematic profiles. The single component fit produces measurements of the emission line velocity and velocity dispersion, as well as the flux for the strong emission lines.

For the gas and stellar velocity maps, spaxels with uncertainties greater than $\pm 15$ km/s were removed from the analysis. For the velocity dispersion maps, following \citet{2017ApJ...835..104V} we removed spaxels where the uncertainty is greater than $10\%$ of the dispersion value plus 25 km/s. 

To assess the degree of rotational support in the SAMI S0s, we employed the widely-used parameter $v/\sigma$, where v is the observed line-of-sight velocity and $\sigma$ is the velocity dispersion. Following recent work with IFU surveys \citep[e.g.][]{2018NatAs...2..483V, 2007MNRAS.379..418C} we calculate $v/\sigma$ using the following equation:

\begin{equation}
\left(\frac{v}{\sigma}\right)^{2} = \frac{\sum_{i=0}^{N_{spx}}{F_{i}V_{i}^{2}}}{\sum_{i=0}^{N_{spx}}{F_{i}\sigma_{i}^{2}}},
\label{refinement_eq}	
\end{equation}

where the sum is over all spaxels within a de-projected radius of 1.5 ${\rm R_{e}}$ and $F_{i}$ is the flux of spaxel $i$. Values of $v/\sigma$ approaching 1 indicate a rotationally supported structure, while values significantly lower than 1 indicate a pressure-supported structure. The de-projected radius of each spaxel corresponds to the major axis of the ellipse on which it is located, which was determined using the $r$-band-derived axis ratio and position angle \citep{2016MNRAS.463..170C}.

The position angles of the gas and stellar components were found by rotating a 3-pixel-wide slit through 360 degrees and adopting the angle which resulted in the maximum velocity gradient.

\begin{figure}
\centering
\includegraphics[width=1.0\columnwidth]{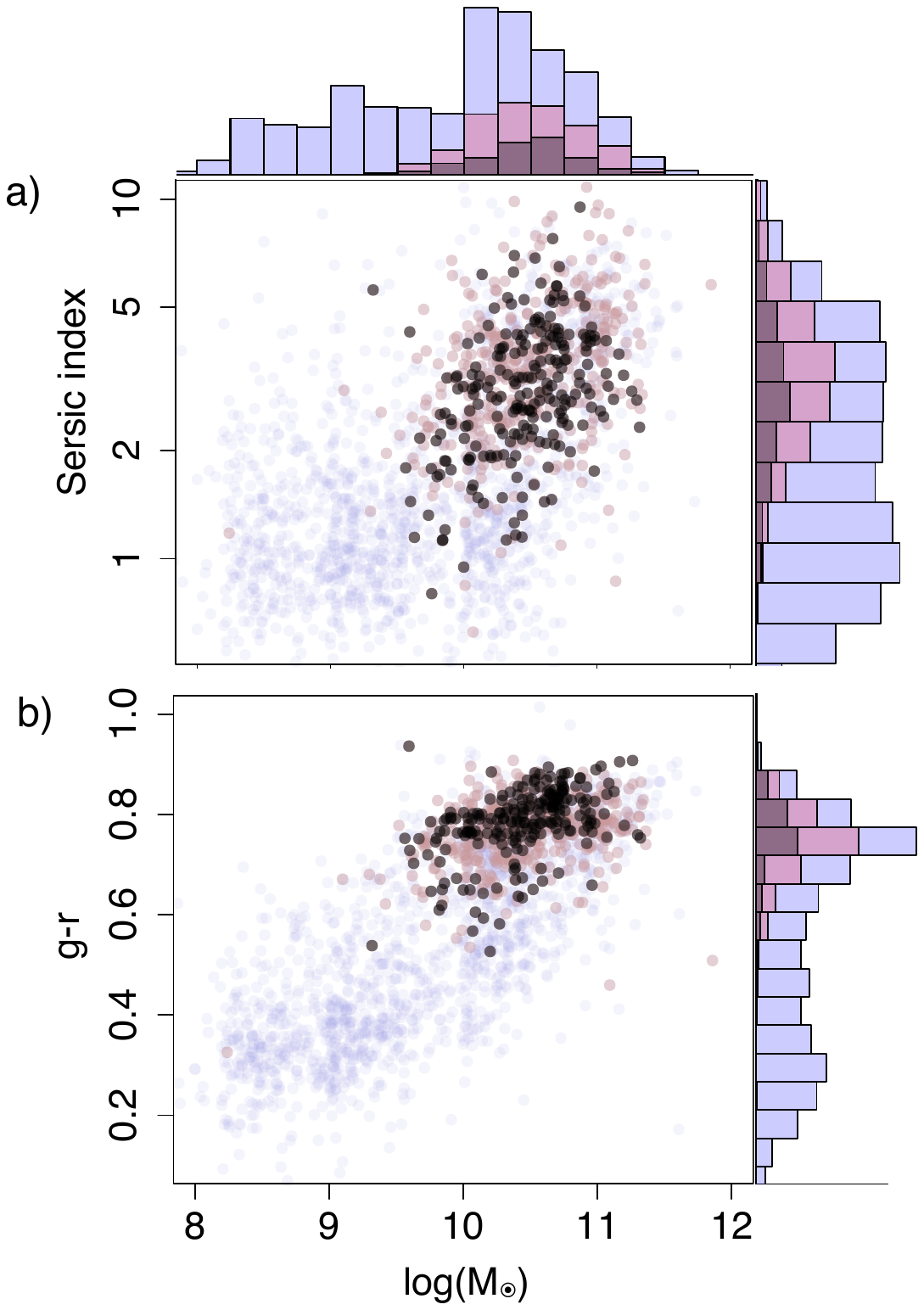}
   \caption{S\'ersic and colour index vs stellar mass for our final S0 sample (black) in comparison to all S0s in the SAMI Survey (red) and the complete SAMI Survey sample (blue) with associated histograms of the distributions on the $x$ and $y$ axis. Our sample selection criteria does not introduce any significant bias relative to the overall S0 population.} \label{bias}
\end{figure}

\subsubsection{Sub-sample}

We initially looked to characterise the range of rotational support observed in our SAMI S0 sample. The range of effective radii covered by the fixed SAMI field of view varies depending on the galaxy's radius and the signal-to-noise ratio of the data away from the bright centre. The number of S0 galaxies within our sample that have useful data beyond a radius of 1.5 ${\rm R_{e}}$ begins to fall rapidly. We therefore selected a cutoff radius of 1.5 ${\rm R_{e}}$; all S0s whose data did not extend out to this radius were removed from the sample. At this distance, the stellar $v/\sigma$ values for faded spiral pathways become more significantly separated from merger pathways due to the decreasing dominance of the central bulge. 

To ensure that the stellar $v/\sigma$ values were not contaminated by the dispersion-dominated centres of the galaxies, all galaxies for which the half width at half maximum of the point spread function was greater than half the effective radius were also removed from the analysis. The remaining sample contained a total of 219 galaxies ('Stellar Kinematics' in Table 1). The distribution of stellar mass vs S\'ersic index and colour for the complete SAMI survey sample (blue), all S0s in the survey (grey) and our S0 final sample (black) is presented in Figure~\ref{bias}, demonstrating that no significant bias is introduced by our sample selection criteria.

\subsection{Radial Profiles}

Following \citep{2016MNRAS.463..170C}, to recover true radial velocity profiles from the line-of-sight velocity projections, we estimated the galaxy's inclination from the measured axial ratio:

\begin{equation}
\cos{i} = \sqrt{\frac{(b/a)^{2}-q_{0}^{2}}{1-q_{0}^{2}}},
    \label{refinement_eq}	
  \end{equation}
 
where $i$ is the inclination, $q_{0}$ is the assumed intrinsic ellipticity, and $a$ and $b$ are the major and minor axis respectively.  We used a commonly-assumed intrinsic ellipticity for a disk galaxy of 0.2 \citep{2000ApJ...533..744T,2016MNRAS.463..170C}. We then applied this inclination to the observed line-of-sight velocity $V_{k\rm{los}}$ of each spaxel $k$ as follows:

\begin{equation}
V_{k} = \frac{V_{k\rm{los}}}{\sin{i}\cos{\theta_{k}}},
    \label{velos}	
  \end{equation}

where $\theta$ is the azimuthal angle of the spaxel $k$ in the galaxy coordinate frame. We determined the direction of rotation of the stellar and gas components by rotating a slice through the 2D map and determining the angle which produced the greatest velocity gradient along that slice. We removed all but the spaxels within 1.5" of the line along the strongest gradient and binned the remaining spaxels by their distance from the galaxy centre. From this, we constructed the radial profile of the velocity, velocity dispersions and $v/\sigma$ using the median values in each bin. 

\subsection{Stellar age profiles}

We use the template-fitting code STARLIGHT \citep{2005MNRAS.358..363C} to determine stellar age profiles for SAMI galaxies. We binned pixels into radial elliptical annuli, with each annulus covering a radial extent of 2 pixels. For each bin, we determined the median flux at each wavelength of all spaxels. We used stellar population templates from the BC03 model \citep{2003MNRAS.344.1000B} for the fitting process. During the fitting process, STARLIGHT determines the mixture of stellar population templates which produces the best match to the observed spectrum. From this template mixture, we determined the final age of each radial bin (along with its metallicity) by finding the weighted average age of templates giving the best match. We estimated uncertainties in the derived stellar ages by using a Monte Carlo method, whereby random noise was added to the median spectrum in proportion to the standard error of the flux at each wavelength in the bin.

\begin{figure}
\includegraphics[width=\columnwidth]{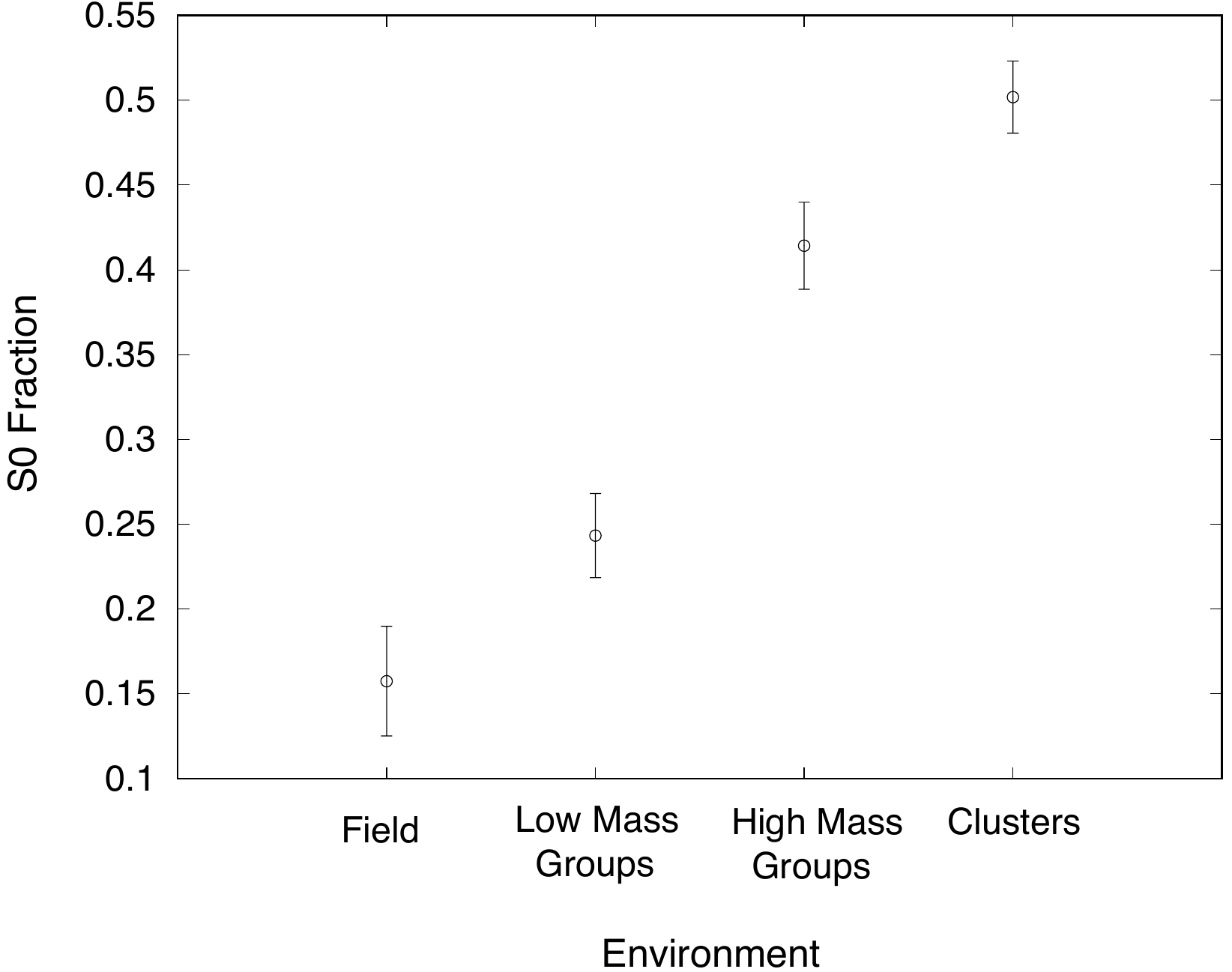}
   \caption{Fraction of galaxies in the field, low (${\rm log(M_{\odot})}<13$) and high ($13<{\rm log(M_{\odot})}>15$) mass groups and clusters which are classed as S0. The error bars denote the 95 per cent bimodal confidence intervals. As expected from previous results, the number of S0 galaxies steadily increases with increasing density.}
 \label{fraction}
\end{figure}

\section{Results}

In this section we first describe the environmental distribution of our S0 galaxy sample. We then characterise the spread in physical properties of these S0s, with a focus on their kinematics and structural parameters. After revealing a wide spread in the stellar $v/\sigma$ values, we show that the optical morphology, gas kinematics and age profiles of S0s at the extreme edges differ between these two groups. 

\label{results}

\begin{table}
\begin{center}
 \caption{Galaxy samples used in this work.}
 \label{data_table}

\begin{tabular}{c|c|c|c|c}

Class&Region&All&Stellar Kinematics&Gas Kinematics\\
(1)&(2)&(3)&(4)&(5) \\
\hline
E&GAMA&104&35&22\\
%\cline{2-5}
&Cluster&100&43&12\\
\hline
S0&GAMA  &354&134&80\\
%\cline{2-5}
&Cluster  &259&85&11\\
\hline
Sp&GAMA&1030&116&98\\
%\cline{2-5}
&Cluster&137&30&9\\
\hline
\end{tabular}
\\[4.5mm]
\end{center}

Classes in column 1 are described in Section~\ref{methods}. Column 2 lists the regions of the SAMI survey, column 3 lists the total numbers of each galaxy class in the SAMI survey, and column 4 lists the number of galaxies in the final sample with a PSF < 0.5 ${\rm R_{e}}$, ellipticity > 0.2 and with stellar kinematic data out to 1.5 ${\rm R_{e}}$. Column 5 shows the number of galaxies which also have gas kinematic data out to 1.5 ${\rm R_{e}}$. 

\end{table}

\begin{figure*}
\includegraphics[width=2\columnwidth]{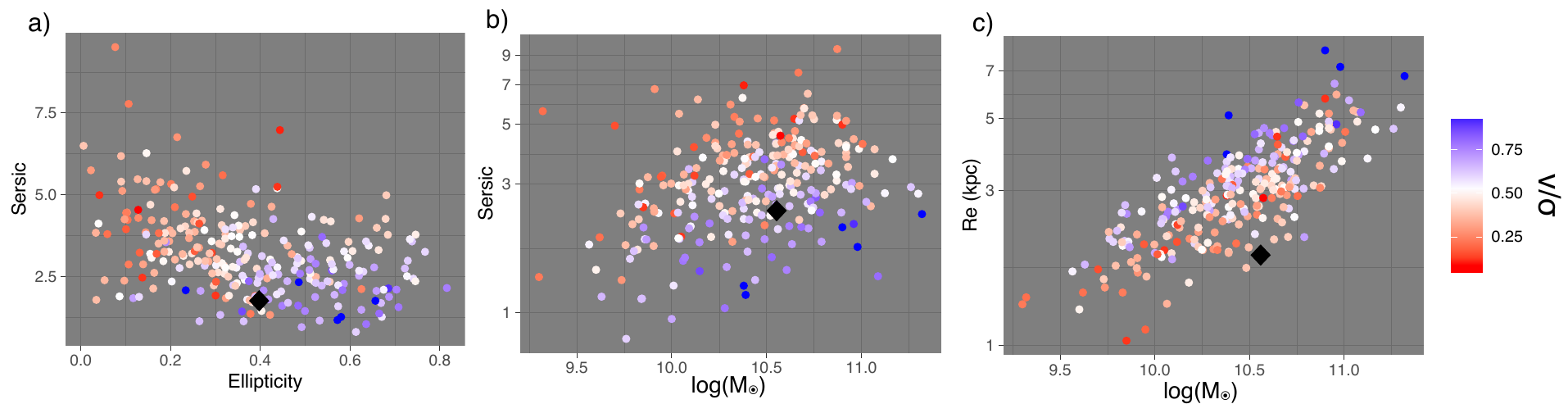}
   \caption{Distributions of our final S0 sample over S\'ersic index vs ellipticity (a), vs stellar mass (b) and the effective radii vs stellar mass (c). Points are shaded according to their $v/\sigma$ value, with bluer points showing more rotationally-supported S0s. A gradient in $v/\sigma$ is evident over each of these parameter spaces, showing that higher rotational support corresponds to lower S\'ersic indexes and larger effective radii. The black diamond corresponds to the locations of the Diaz et al. (2018) fiducial model, which has a stellar $v/\sigma$ value of around 0.2. }
 \label{clustering}
\end{figure*}

\begin{figure}
\centering
\includegraphics[width=0.8 \columnwidth]{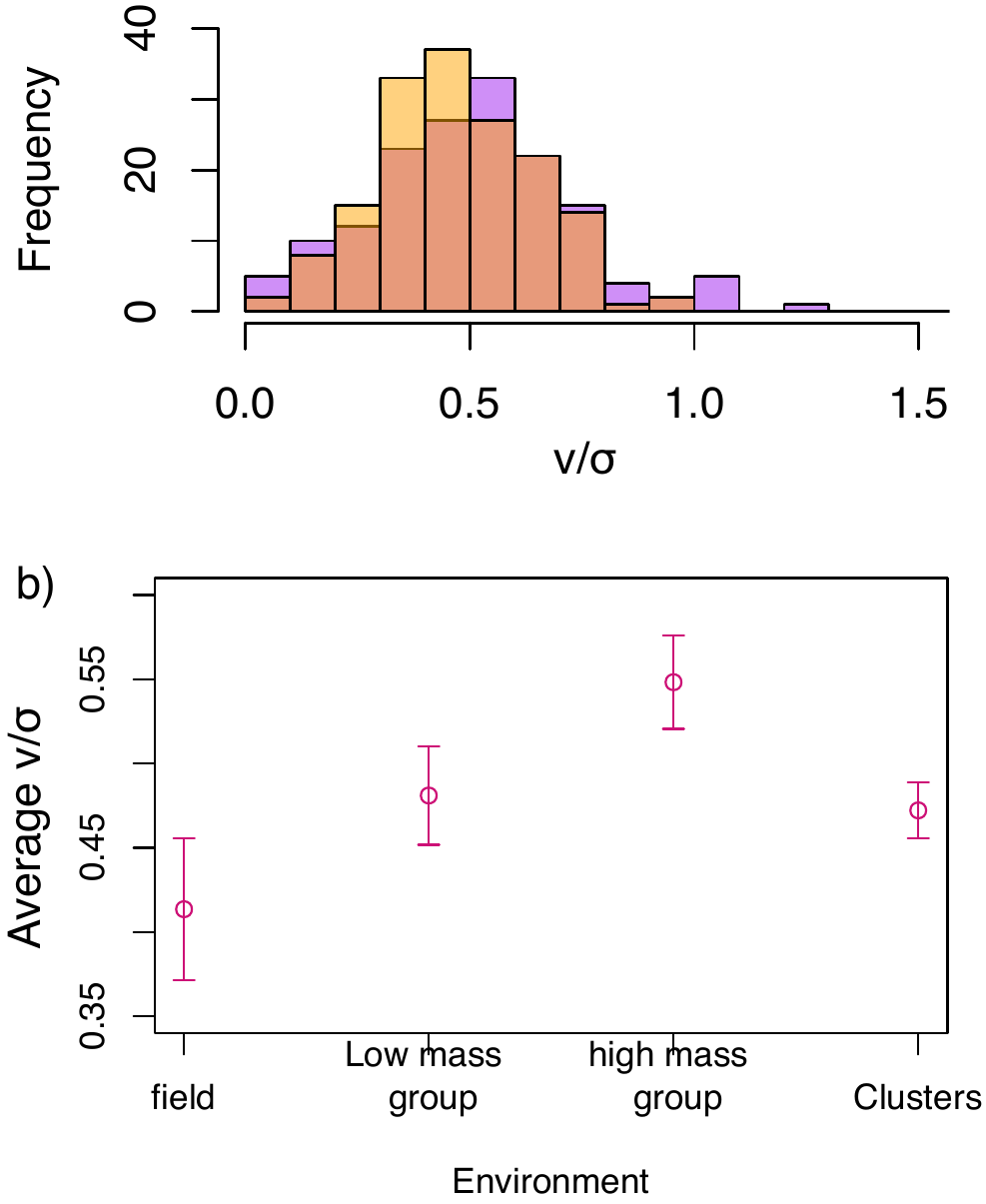}
   \caption{Distribution of $v/\sigma$ values (a) for the GAMA (purple) and cluster (orange) regions, and the average stellar $v/\sigma$ of S0s in the field, low mass groups (below $10^{13}M_{\odot}$, high mass groups (above $10^{13} M_{\odot}$) and the cluster regions (b). } 
\label{env}
\end{figure}

\subsection{Host environments of S0s}

We first looked at the distribution of S0s across the different environments covered by SAMI. The fraction of all galaxies in the SAMI survey classified as S0s is shown in Figure~\ref{fraction} as a function of environment. Here the full galaxy sample is used ('All' in Table 1). As the mass of the host group increases, there is a strong increase in the fraction of S0 galaxies, consistent with previous findings \citep[e.g. ][]{1980ApJ...236..351D}. S0 galaxies account for around half of all galaxies located in the cluster regions, while within the field and low mass group environments, S0 galaxies still make up a non-negligible fraction of 0.16 and 0.24 respectively. This also demonstrates that our SAMI sample contains a significant fraction of S0s across all environments including the field and low mass groups. 

\subsection{Stellar Kinematics and S0 Subpopulations }
 
\begin{figure*}
\includegraphics[width=1.7\columnwidth]{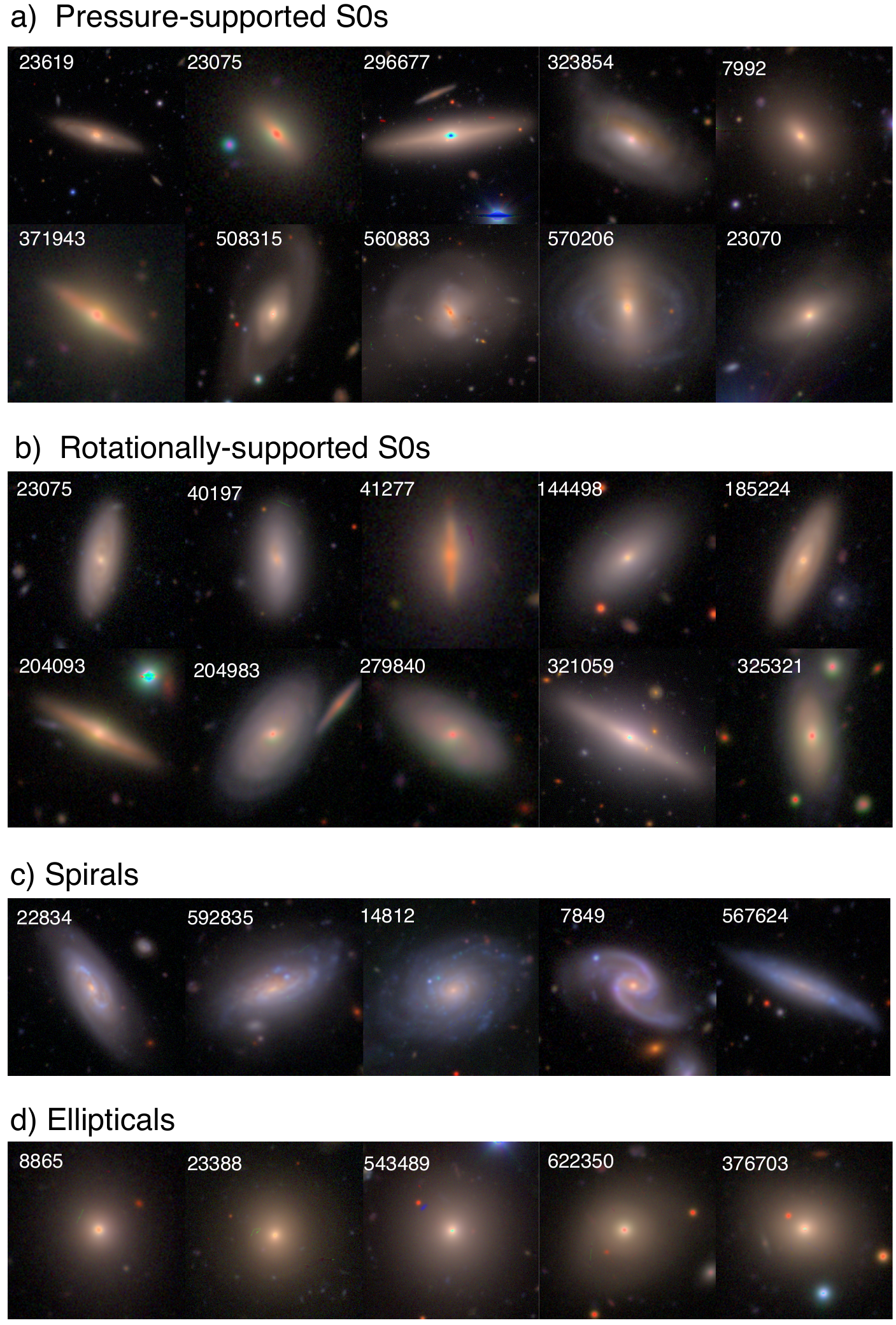}
   \caption{HSC imagery of S0 galaxies in the pressure-supported population ($v/\sigma$ < 0.5, a) and rotationally-supported population ($v/\sigma$ > 0.5, b). Many pressure-supported S0s feature signs of disruption such as shells and tidal arms (e.g. 323854), while those in the rotationally-supported group mostly feature smooth well-defined disks, some of which show signs of remanent spiral structures. For comparison, examples of HSC imagery of spirals and ellipticals are shown in c) and d) respectively. The SAMI catalogue ID number of each galaxy is shown in the top left of each image}
 \label{HSC}
\end{figure*}

We next looked to characterise the range of physical properties of S0s within the SAMI sample. A large range, or any bimodality present in the distributions, may indicate a wide range of conditions in the formation of S0s, or indeed the presence of multiple formation pathways. We focus here on the S\'ersic index, effective radii and stellar $v/\sigma$, since these properties together physically characterise the structure of the S0s and are affected by both the initial transformation and subsequent evolution. Figure~\ref{clustering} presents the distribution of the final S0 sample (including both the GAMA and cluster regions) in the S\'ersic vs ellipticity, S\'ersic vs galaxy mass and effective radius vs stellar mass planes, with the points coloured by their stellar $v/\sigma$ values. The distributions of S0s over each of the parameters, and across each of the parameter spaces shown, appears to be continuous with no clear separation into bimodal populations. 

It is however evident that there is a wide range in rotational support within the S0 population. In particular, the range in $v/\sigma$ over the different parameter spaces forms a strong gradient in each case. S0s with low rotational support ($v/\sigma$ below 0.5) generally feature lower ellipticities, higher S\'ersic indices, slightly higher stellar masses and smaller effective radii. These features indicate a more compact system with a higher degree of random stellar motions. The more rotationally-supported S0s ($v/\sigma$ above 0.5), meanwhile, tend to feature larger ellipticities, lower S\'ersic indices, and larger effective radii. These taken together indicate that they feature a more prominent rotationally supported disc component. While the correlation between $v/\sigma$ and ellipticity will be partly due to projection affects, the additional correlations between other structural parameters indicate that there is also an intrinsic shift in the structure of the S0s from high to low measured $v/\sigma$ values. The lower ellipticites in the pressure-supported population may therefore also indicate an overall shift to more spherical structures, supported by the larger S\'ersic index and higher degree of random motion.

Figure~\ref{env} shows the average stellar $v/\sigma$ values for S0s in the field, low mass (up to $10^{13}M_{\odot}$, high mass (over $10^{13} M_{\odot}$ and cluster regions. The observed trend in the groups is marginal relative to the uncertainty range, with S0s in the field featuring the lowest average stellar $v/\sigma$, increasing towards the high-mass groups, and then falling back in the cluster region. This reversal in the clusters could be due to either an increase in harassment experienced by these galaxies or the merging of group and field S0s into the clusters; this is discussed in Section~\ref{environment}. Alternatively, this may be a bias resulting from different sample selections between the GAMA and Cluster regions.

\begin{figure}
\centering
\includegraphics[width=0.9\columnwidth]{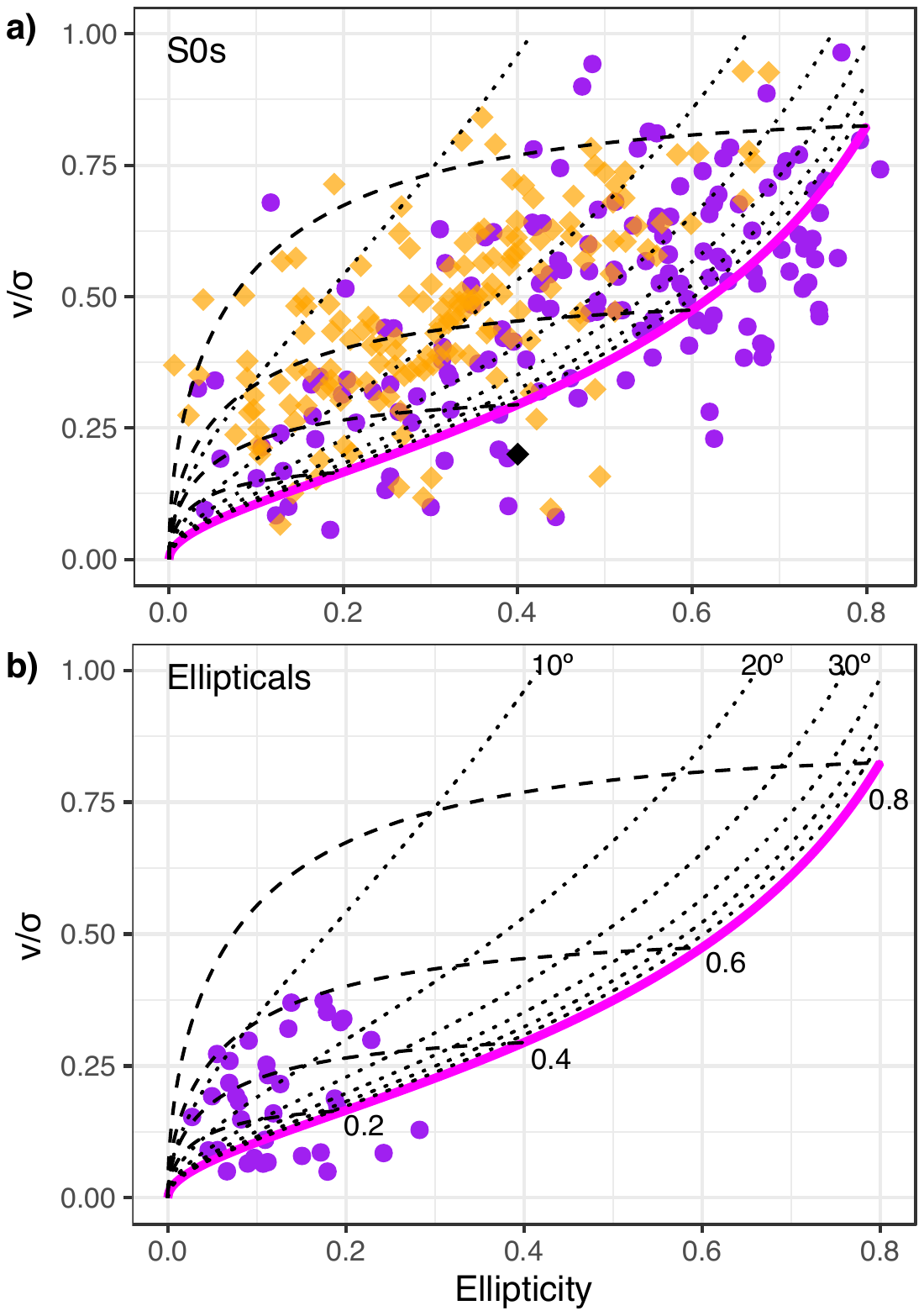}
   \caption{a) $v/\sigma$ vs ellipticity for S0s in the GAMA (purple circles) and cluster (orange diamonds) regions. The magenta line shows the expected values for an edge-on axisymmetric galaxy with anisotropy $\delta = 0.7 {\rm \epsilon_{intrinsic}}$. The dashed lines show this relation for different inclinations, while the dotted lines show positions of galaxies with equal intrinsic ellipticity \citep{2007MNRAS.379..418C}; the respective values for each line drawn is displayed in b). The black diamond shows the location of the Diaz et al. (2018) fiducial model. For comparison, b) shows the elliptical population in the GAMA region, illustrating the previously identified slow and fast rotators in relation to the range of $v/\sigma$ seen amongst the S0s.}
\label{ellip_kin}
\end{figure}

Hyper Suprime Cam (HSC) images of S0s with stellar $v/\sigma$ above and below 0.5 are shown in Figure~\ref{HSC}. These high-resolution images are currently only available for the GAMA region. The galaxies shown here were blindly selected from a list of randomised catalogue IDs. More rotationally supported S0s generally feature smooth, well-defined disks with hints of remnant spiral structure in some cases, suggesting a passive evolutionary history. Many S0s in the pressure-supported regime (for example 323854 and 508315) feature signs of interactions and mergers, including shells, tidal features and structures with orientations different to that of the rest of the galaxy. 

Figure~\ref{ellip_kin} a) shows the distribution of our S0 sample on the widely-used stellar $v/\sigma$ vs ellipticity plot for the GAMA and cluster region. For comparison, Figure~\ref{ellip_kin} b) shows the same distribution for ellipticals in the GAMA region which meet the same kinematic selection criteria, with the slow rotators highlighted in orange. This illustrates that the distribution in kinematics seen among the S0s here encompasses a wider range than that covered by the slow and fast-rotating ellipticals. The cluster S0s appear to be shifted towards lower ellipticities relative to the GAMA region. As a check for whether this is caused by a systematic bias in ellipticity measurements between the two regions due to different imagery (SDSS images for the GAMA region and VST images for the cluster regions, see Section~\ref{environment_methof}), we also looked at the distribution of ellipticity for spiral galaxies and found no such shift. In addition, where an overlap between SDSS and VST was present, parameters derived from the two sets of imagery were checked for consistency \citep{2017MNRAS.468.1824O}. This suggests that S0s in the cluster region have intrinsically lower ellipticities, which could be due to the increased harassment these galaxies experience in the denser environment. 

As a comparison, Figure~\ref{ellip_kin} b) also illustrates the stellar $v/\sigma$ vs ellipticity distribution of the elliptical population within GAMA. This demonstrates that the well-known separation of centrally slow and fast rotator ellipticals occurs outside the large range of rotational support seen amongst the S0s \citep{2016ARA&A..54..597C}.

\subsection{Gas Kinematics}

\begin{figure}
\includegraphics[width=0.7\columnwidth]{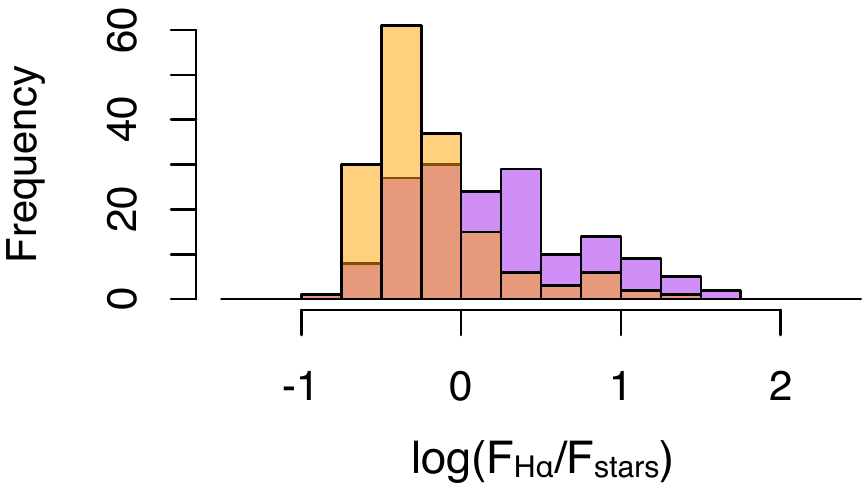}
   \caption{Distribution of the integrated {\rm $H\alpha$} flux to underlying continuum stellar flux ratio for S0s in the GAMA (purple) and cluster (orange) S0 regions. S0s in the clusters feature less gas emission than those in the rotationally-supported group.}
 \label{gas_distribution}
\end{figure}

\begin{figure}
\centering
\includegraphics[width=1.0\columnwidth]{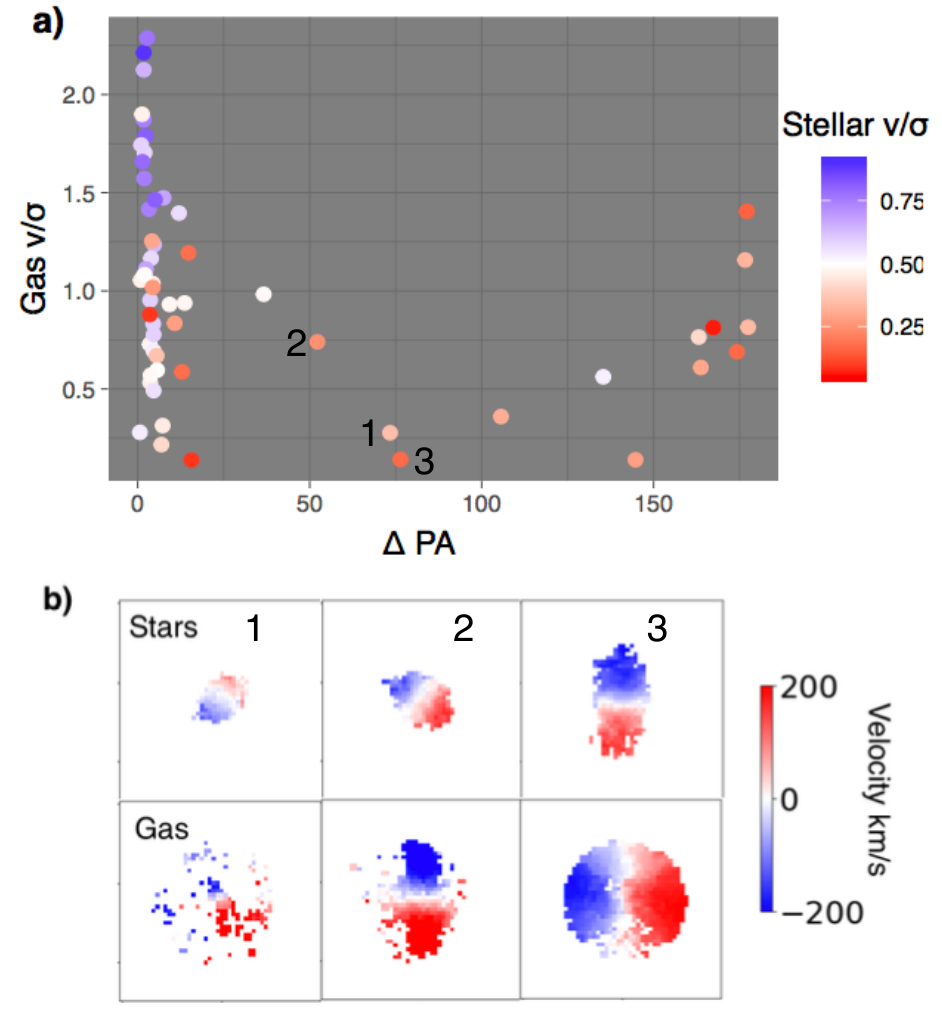}
   \caption{a) Distribution of the stellar and gas $v/\sigma$ as well as the degree of misalignment between the gas and stellar components in the S0 sample with gas kinematics available, again coloured by their $v/\sigma$ values. S0s with high stellar $v/\sigma$ values all have aligned gas components, while those with misaligned gas and stellar components feature $v/\sigma$ values below 0.5. b) Examples of gas and stellar kinematic maps for S0s with dPAs close to 90 degrees, illustrating that the measured PA differences are not due to poor kinematic maps.}  \label{including_gas}
\end{figure}

\begin{figure*}
\centering
\includegraphics[width=1.8\columnwidth]{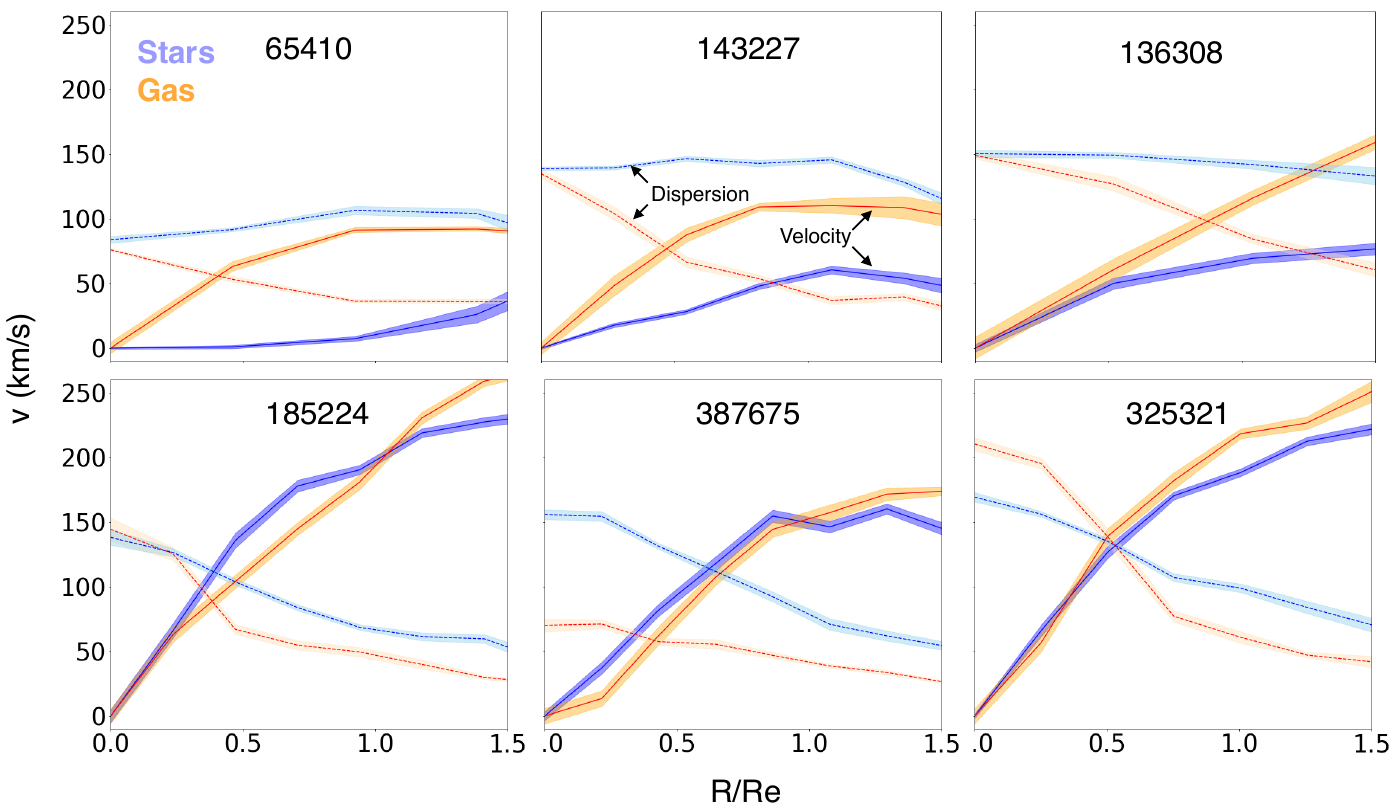}
   \caption{Stellar (blue) and gas (orange) kinematic profiles for S0s with $v/\sigma$ below 0.5 (top row) and above 0.5 (bottom row). Solid lines are the velocities, dotted lines are the velocity dispersions. The pressure-dominated S0s feature gas rotation velocities significantly higher (at least two times faster at 1.5 Re) than the stellar components, while the rotation-dominated S0s feature gas velocities in line with their stellar components.} 
 \label{kinematic_profiles}
\end{figure*}

To gain further insight into the structure and history of the S0 sample, we next investigated the distribution of gas kinematics within the S0s and determine how these relate to the stellar component. Not all S0s in our sample have line emission strong enough to derive accurate gas kinematics (see Table 1), as would be expected given the typically low gas contents of this class of galaxy. This is particularly true for the cluster regions, where both the larger redshifts and the generally lower gas contents of galaxies in cluster environments would result in more S0s falling below the emission detection threshold. The distribution of the {\rm $H\alpha$} flux weighted by the stellar flux for both the GAMA and cluster region is shown in Figure~\ref{gas_distribution}. The distribution for S0s in the GAMA region (i.e. those in field and group environments) is broader with a tail towards higher {\rm $H\alpha$} flux, while the majority of cluster S0s contain very little detected gas emission. Because of this, we cannot derive accurate gas kinematics for many S0s in the cluster region.

For S0 galaxies meeting the additional selection criteria for inclusion in the analysis of gas kinematics, Figure~\ref{including_gas} shows the distributions of the stellar and gas $v/\sigma$, as well as the difference in position angle between the stellar and gas components, again coloured by their stellar kinematics. S0s with unusual PAs were visually investigated to confirm that these differences are not due to poor kinematic maps; three examples are shown in Figure~\ref{including_gas} b). Most evident here is that the S0s with highly rotationally-supported star kinematics also feature rotationally-supported gas kinematics, as well as very small differences in the position angles of the two components. This indicates that the stellar and gas components in these galaxies are co-rotating in a common structure. S0s with $v/\sigma$ values below 0.5, however, feature a large range of misalignments between the stellar and gas components. The six S0s with PAs between 50 and 150 degrees have lower $v/\sigma$ values relative to those near 0 or 180 degrees; such configurations are likely more stable in systems where the stellar component has a lower degree of angular momentum. Misalignments of S0 star and gas components in SAMI have already been noted \citep{2019MNRAS.483..458B}; here we are demonstrating that this misalignment only occurs in S0s with low rotational support. 

Figure~\ref{kinematic_profiles} shows typical examples of kinematic profiles for S0s in the pressure-supported S0s (top row) and rotationally-supported S0s (bottom row). In the top row, the gas components are rotating significantly faster than the stellar components, while in the S0s from the rotationally-supported group, the rotational velocities of the gas and stellar components are similar. This further suggests that the star and gas components in the rotationally-supported S0s have evolved together, while the gas component in the pressure-supported group may have a separate origin to the stellar component. However, we note that S0s with low stellar $v/\sigma$ may also be more likely to feature mis-aligned gas kinematics simply due to a larger bulge to disk ratio. 

\subsection{Age Profiles}

\begin{figure}
\centering
\includegraphics[width=0.8 \columnwidth]{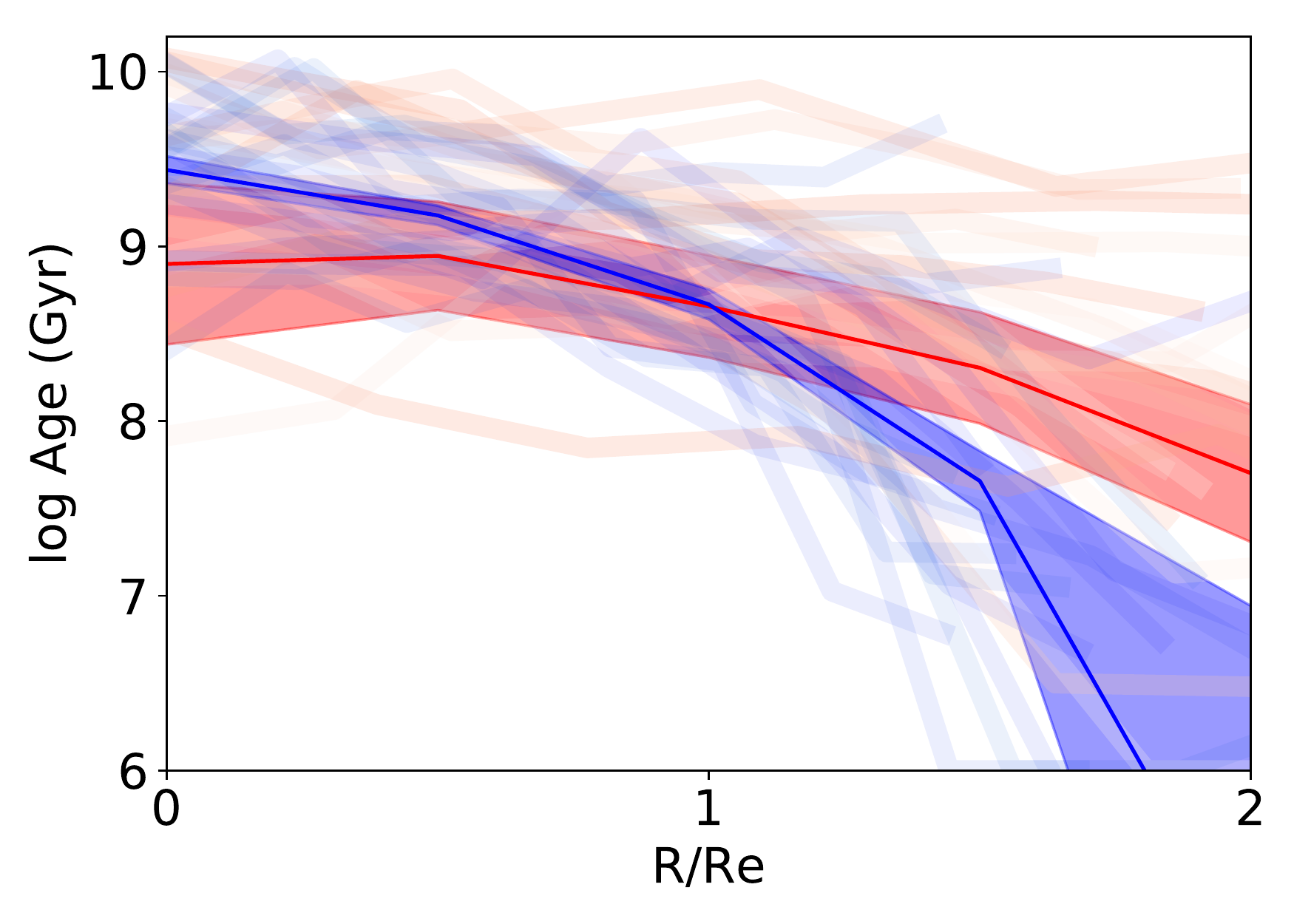}
   \caption{Stellar age profiles of pressure and rotationally-supported S0 galaxies in the GAMA region. The faint lines are the profiles of individual S0s with $v/\sigma$ > 0.75 (blue) or < 0.25 (red), while the solid line and shading are the average profile and its uncertainty respectively. Age estimates were not reliable at larger radii for some S0s, indicated by the truncated lines of some of the individual profiles. While the profiles are largely consistent, the rotational S0s show signs of a rapid age drop towards the outer regions of the galaxy.} 
\label{age_profiles}
\end{figure}

Finally, we examine whether there is any difference in the stellar age profiles of S0s at the extreme ends of the stellar $v/\sigma$ distribution. Figure~\ref{age_profiles} shows the age profiles of pressure-supported and rotationally-supported S0s, along with the average profile for each. For this comparison, we use the limits $v/\sigma$ > 0.75 for the rotationally-supported group (blue) and $v/\sigma$ < 0.25 for the pressure supported group (red), resulting in 19 and 20 S0s respectively. 

The average age profiles have central ages consistent with each other within the uncertainties. The rotationally-dominated S0s do show signs of a rapid drop in age towards the outer regions, suggesting that the more prominent disk structure in these galaxies consists of younger stars. However this difference relative to the pressure-supported S0s is not significant within this sample. 

\section{Discussion}

In this section we firstly compare our findings with the previously-known distinction between centrally slow and fast rotating early-type galaxies. We then discuss whether the range of properties seen in the S0 population are the result of different formation pathways. As described previously, the two main formation pathways proposed involve either the fading of spirals to S0s, or disruptive minor merger events. The former is expected to lead to S0s which retain the disk structure and a significant amount of the rotational support of the former spiral, while the latter process would increase the degree of pressure support and shift the stellar population to a more centrally-concentrated structure. While we observe a continuum rather than any clear bimodal separation in the stellar kinematics, variations in the specific physical parameters of each occurrence of an S0 formation process ( for example different mass ratios and relative angular momentum in minor mergers) would lead to a range in values of the final degree of rotational support, blurring the separation between the two populations. 

\subsection{Relation to the slow vs fast rotator classification of early-type galaxies}
\label{discuss_slow_fast}

A large amount of work has recently been done on the separation of early type galaxies (ellipticals and S0s) into slow and fast rotators. Slow rotators are defined to be those galaxies which fall within the region below $\lambda_{\rm R_{e}}=0.08+e/4$ \citep{2016ARA&A..54..597C}. Recent work has suggested that slow and fast rotators may have different evolutionary pathways, with the slow rotators resulting from merger activity and the fast rotators following a more passive evolution \citep{2017MNRAS.468.3883P}. However, the majority of early-type galaxies which fall within the slow-rotator class are ellipticals; for example in the work of \citet{2007MNRAS.379..401E}, while 10 out of 25 ellipticals are classed as slow rotators, only two out of 22 S0s are classed as slow rotators while the remaining were classed as fast rotators. Here we found the same behaviour for our S0 sample; only 12 out of 219 S0s lie within the slow-rotator region. This highlights that the spread in kinematics of S0s we find here lies in a different region of the ellipticity vs $\lambda_{\rm R_{e}}$ parameter space. Therefore we conclude that the spread in $v/\sigma$ we observe is not caused by subpopulations corresponding to the slow and fast rotators, but is caused by a range of processes intrinsic within the S0 class. However, it should be noted that kinematics at larger radii considered here can differ from the kinematics derived from smaller radii, as demonstrated in \citet{2016MNRAS.457..147F}.

\subsection{Origin of pressure-supported S0s}

The more pressure-supported population of S0s (those with $v/\sigma$ less than 0.5) feature a pressure-supported stellar component, high S\'ersic indices, lower ellipticites, and are more prominent in lower density environments. Many of these S0s show signs of disruption in their optical imagery, such as shells and tidal streams. Their gas content is lower than that of the S0s with a higher degree of rotational support, and the position angle of the gas rotation relative to the stellar component covers a large range from aligned to anti-aligned, suggesting that the gas component has a different origin to the stellar component. This is in contrast to spiral galaxies, for which the vast majority feature aligned gas and stellar components \citep{2019MNRAS.483..458B}. 

\begin{figure*}
\centering
\includegraphics[width=1.4\columnwidth]{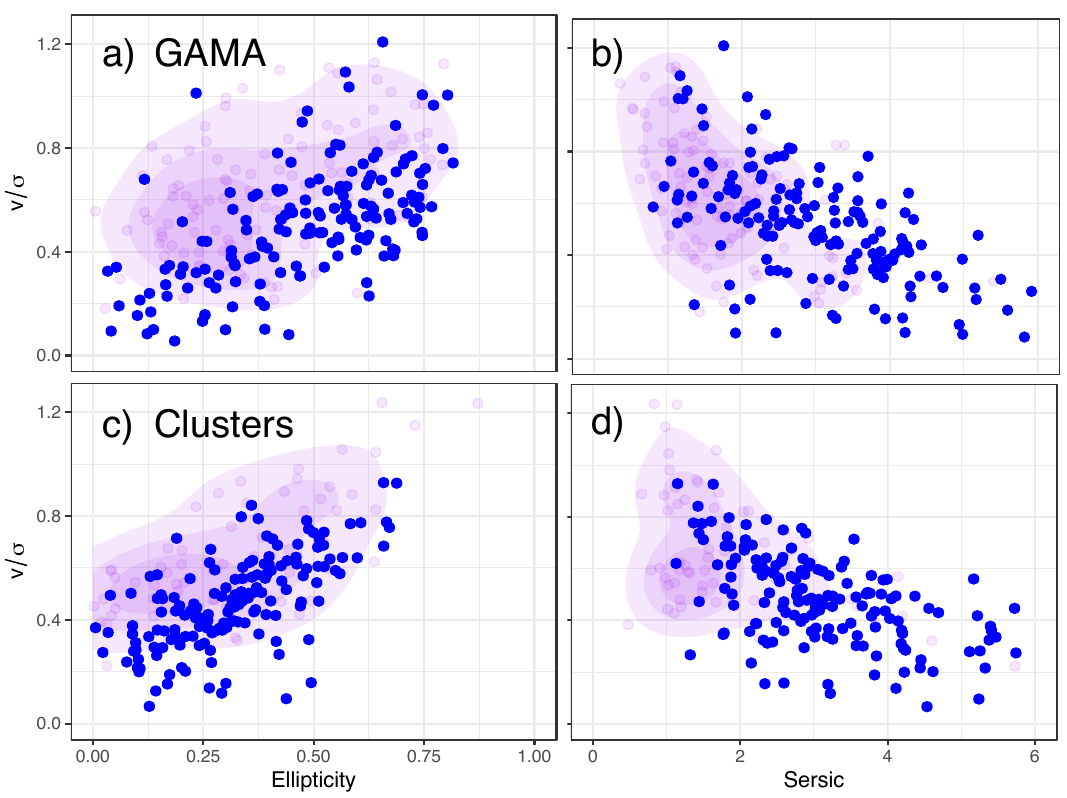}
   \caption{Comparison between the S0 population in this study and a mass-matched sample of spiral galaxies (purple points and contours) in the $v/\sigma$ vs ellipticity (a,c) and S\'ersic index (b,d) in the GAMA region (top row) and cluster regions (bottom row). The more rotationally-supported S0s are located within the same parameter-space regions as the spiral population, while the pressure-supported S0s with higher S\'ersic indices lie outside the regions of the spiral population.}   
 \label{comp_w_spirals}
\end{figure*}

S0s forming via minor mergers or tidal interactions are expected to feature a higher degree of pressure support \citep{2018MNRAS.476.2137R}, consistent with what is seen in these S0s. Comparing the pressure-supported S0s with spiral galaxies in the SAMI survey (Figure~\ref{comp_w_spirals}) shows that the pressure-supported S0s feature higher S\'ersic indices and higher stellar masses than the SAMI spiral population. If they descended directly from spirals, their progenitors would have been a subpopulation of spirals with high masses and a dominating bulge. However, their $v/\sigma$ values are lower than spirals of similar mass, making them difficult to explain with a passive faded-spiral pathway. 
 
 All these features could indicate that these S0s have formed via a disruptive process such as a minor merger or tidal interaction, rather than via the fading spiral pathway. However, it cannot be ruled out that these are older systems which have experienced a greater amount of harassment, ram pressure stripping or significant disruption since their initial formation. 

Alternatively, as seen in \citet{2019MNRAS.485.2656C}, passive galaxies may also feature lower v/$\sigma$ values due to a slow-down in angular momentum build-up once they enter a denser environment. S0s with lower v/$\sigma$ values seen in the larger group environments may therefore also reflect a passive evolution rather than a merger event. Comparisons with cosmological simulations may help to determine whether an altered passive evolution or a more disruptive merger event is resulting in S0s with a greater degree of pressure support in these environments.

\subsection{Are pressure-supported S0s consistent with a cE+disk origin?}

\begin{figure*}
\includegraphics[width=1.9\columnwidth]{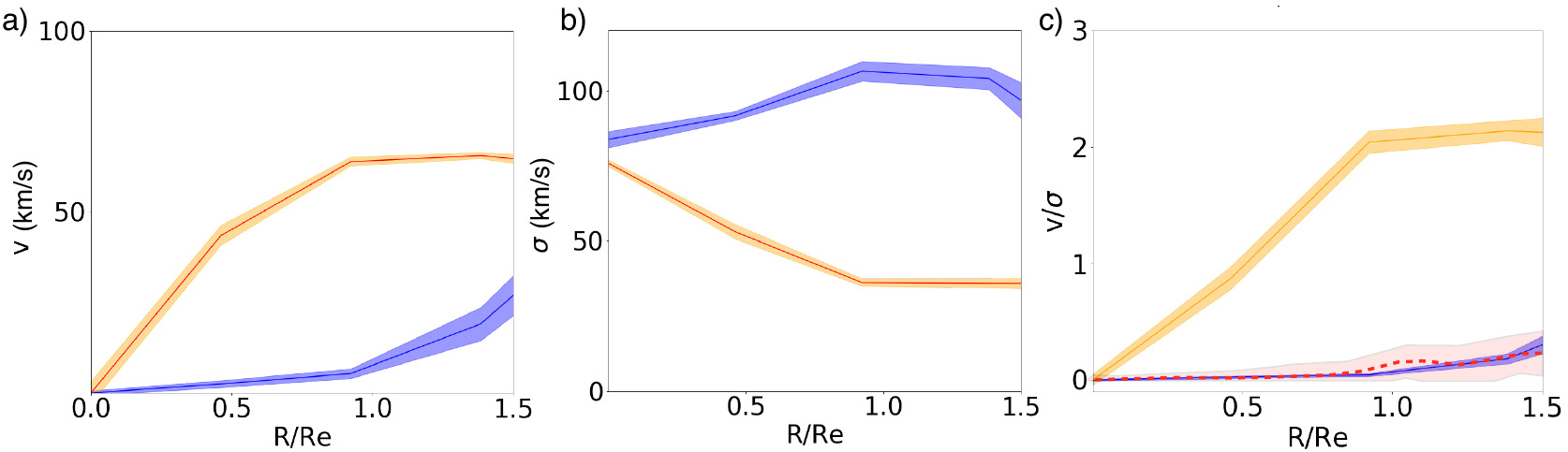}
   \caption{Example of an S0 galaxy with kinematics similar to the Diaz et al. (2018) model. The prediction of the stellar $v/\sigma$ profile from Diaz et al. is overlaid on c) as a red dashed line and shaded region denoting the standard deviation. a) shows the rotational velocity of the stellar (blue) and gas (orange) component. Shaded regions delimits one standard error. The gas component is rotating significantly faster than the stellar component, as expected if the gas is located in a tight disk. b) shows the velocity dispersion profiles, and c) shows the $v/\sigma$ profile, with the Diaz et al. (2018) fiducial model overlaid (red line and shaded region). The standard error of the mean values in each radial bin is indicated by the shaded regions.}
 \label{65410}
\end{figure*}

The comparison of our observational results to the fiducial model in Diaz et al. (2018) are consistent with the claim that at least some S0s may have resulted from the cE+disk merger pathway. The Diaz et al. (2018) simulations predicted that the cE+disk merger would lead to an S0 with low stellar $v/\sigma$, a faster-rotating gas component and a negative age gradient. The final S0 in the fiducial model has a mass of around $10^{10.5}{\rm M}_{\odot}$. 

Its degree of rotational support lies among the pressure-supported S0s; the position of the model S0 on the ellipticity vs  $v/\sigma$ diagram is shown as the black diamond in Figure~\ref{ellip_kin}. Figure~\ref{clustering} also displays the locations of the fiducial model within the S\'ersic vs ellipticity, S\'ersic vs galaxy mass and effective radius vs stellar mass planes relative to the observed S0 distributions. An example of an S0 galaxy lying near this point in ellipticity vs $v/\sigma$ space is shown in Figure~\ref{65410}. Consistent with the fiducial model, the stellar $v/\sigma$ remains below 0.5 out to 1.5 effective radii, while the gas component rotates significantly faster.  

The kinematic predictions presented in Diaz et al. (2018) corresponded to only a single scenario. It is certainly possible that a range of masses and properties of the cE and the disk galaxy, along with their relative motions, would lead to S0s with a range of masses and kinematic properties. For example, a more massive disk with a faster relative velocity may lead to a more rotationally supported S0 with a relatively larger disk component and higher $v/\sigma$ values. Therefore, the fraction of the S0 population consistent with the model could be significantly higher than what can be determined at this stage.

\subsection{Origins of the rotationally-supported ($v/\sigma$ > 0.5) S0s}

The more rotationally-supported S0 galaxies are located more frequently in higher density groups. The majority of these S0s feature smooth, well-defined disks, some of which show hints of remnant spiral features. The gas in these S0s is also co-rotating with the stellar component in the majority of cases, which suggests that the gas has co-evolved with the stellar component. 

A spiral galaxy which fades to an S0 would be expected to maintain a constant mass, since no significant amount of new mass enters the galaxy through a merger \citep{2018MNRAS.476.2137R}. In addition, the lack of any significant disruptions would leave the concentration of the stellar component similar to the original spiral, and the galaxy would be expected to maintain its degree of rotational support. We would therefore expect a faded spiral to have similar mass, $v/\sigma$ and S\'ersic indices to spiral galaxies. Comparing the rotationally-supported S0s to the mass-matched sample of spiral galaxies (Figure~\ref{comp_w_spirals}) shows that in both the GAMA and cluster regions, these S0s feature similar concentrations and kinematics to the spiral population, which lends support to the idea that these S0s form via the faded spiral pathway. The shift to higher S\'ersic and lower $v/\sigma$ values from the GAMA to the cluster regions evident in Figure~\ref{comp_w_spirals} is likely a reflection of the morphology-density relation, where the higher-density cluster environment contains a higher proportion of bulge-dominated systems.

 The mass distribution of the full SAMI spiral population ranges from $10^{8}$ to $10^{11} M_{\odot}$; the lack of rotationally-supported S0s with lower masses may indicate that only high-mass spirals undergo this process. In addition, barring any major subsequent inflow of gas, S0s originating from faded spirals would retain the co-rotation of the stellar and gas component seen in spiral galaxies, which is what we also observe in the rotationally-supported S0 population. 

\subsection{Environment}
\label{environment}
Due to the higher relative velocities between galaxies in higher mass groups and clusters, the occurrence of mergers and slow gravitational interactions is expected to be less than that in lower-density environments \citep{1992ARA&A..30..705B}. Meanwhile, the increased density of the intra-cluster medium in higher-mass systems may lead to the increased occurrence of fading pathways such as ram pressure stripping. The increase in $v/\sigma$ observed in higher mass groups (see Figure~\ref{env}) could therefore reflect a decreasing occurrence of the merger formation pathway and an increase in the prevalence of the fading-spiral transformation pathway. 

Interestingly, the decrease in the average $v/\sigma$ seen in the cluster regions goes against the trend seen in the groups (Figure~\ref{env}). This decrease could be the result of S0s which have formed within small groups merging into the clusters. Indeed, \citet{2010ApJ...711..192J} showed that S0 formation is occurring more prominently in low-mass groups, and that the relative numbers of S0s seen in clusters can be explained by the subsequent merging of these low-mass groups into the clusters. Alternatively, additional harassment in the cluster environment, such as multiple flybys of other cluster members, may lead to further heating of the disk. Future work will look at the morphology-density relation of S0s within clusters, in particular their stellar $v/\sigma$ as a function of radius. 

 \section{Summary and Conclusion.}
 
The aims of this work were to firstly investigate whether there is evidence for subpopulations of S0s with different formation pathways, and secondly to test the cE + disk formation pathway proposed in \citet{2018MNRAS.tmp..724D}. To achieve this, we used the SAMI galaxy survey which contains resolved kinematic data on a large sample of S0s across the full range of environments in which they are found. 
 
 We found a wide range of rotational support, illustrated by the range of stellar $v/\sigma$ within 1.5 effective radii. The variation seen in $v/\sigma$ corresponds to accompanying changes in other structural parameters, with more rotationally supported S0s featuring higher ellipticities, lower spiral-like S\'ersic indices and larger effective radii. This indicates that more rotationally supported S0s feature a more spiral-like structure with a prominent disc, while the more pressure supported S0s feature a more compact spherical structure. 
  
The more pressure-supported (stellar $v/\sigma$ < 0.5) S0s dominate in low-density environments while the more rotationally-supported S0s become more prominent in higher density environments. Optical images of S0s at each end of the $v/\sigma$ range revealed that while both rotationally and pressure supported S0s include those with a bulge and a smooth, well-defined disk, a significant fraction of S0s with $v/\sigma$ > 0.5 show signs of merger or gravitational interactions such as shells and tidal features. 

Comparisons with the Diaz et al (2018) fiducial S0 show that the cE + disk formation model is consistent with S0s in the more pressure-supported regime, and is therefore a viable S0 formation mechanism. Further simulations exploring a wider range of initial conditions are needed in order to determine what fraction of the S0 population could be explained by this model. 

For S0s which also have gas kinematics, we found that all S0s with $v/\sigma$ > 0.5 featured co-rotating stellar and gas components while many S0s with $v/\sigma$ < 0.5 feature a mis-aligned gas component, suggesting that the gas in these galaxies may have an external origin. Finally, age profiles show that the pressure-supported S0s have, an average, slightly flatter age profiles relative to the rotationally-supported group, however the two populations are consistent within uncertainties.
 
 Our observations have identified a wide range of rotational support among S0 galaxies. The corresponding changes in other physical parameters suggest that the rotationally-supported S0s formed by the fading of spiral galaxies, while more complex formation pathways (e.g., mergers, interactions) are likely needed to explain the pressure-supported S0s, which are preferentially observed in the field and low-mass group halos.

\section*{Acknowledgements}
The SAMI Galaxy Survey is based on observations made at the Anglo-Australian Telescope. The SAMI was developed jointly by the University of Sydney and the Australian Astronomical Observatory. The SAMI input catalogue is based on data taken from the Sloan Digital Sky Survey, the GAMA Survey, and the VST ATLAS Survey. The SAMI Galaxy Survey is supported by the Australian Research Council Centre of Excellence for All Sky Astrophysics in 3 Dimensions (ASTRO 3D), through project number CE170100013, the Australian Research Council Centre of Excellence for All-sky Astrophysics (CAASTRO), through project number CE110001020, and other participating institutions. The SAMI Galaxy Survey website is http://sami-survey.org/.
 
GAMA is a joint European-Australasian project based around a spectroscopic campaign using the Anglo-Australian Telescope. The   GAMA input catalogue is based on data taken from the Sloan Digital Sky Survey and the UKIRT Infrared Deep Sky Survey. Complementary imaging of the GAMA regions is being obtained by a number of independent survey programmes including GALEX MIS, VST KiDS, VISTA VIKING, WISE, Herschel ATLAS, GMRT, and ASKAP providing ultraviolet to radio coverage. GAMA is funded by the STFC (UK), the ARC (Australia), the AAO, and the participating institutions. The GAMA website is http://www.gama-survey.org/
 
This work was supported through the Australian Research Council's Discovery Projects funding scheme (DP170102344). Support for AMM is provided by NASA through Hubble Fellowship grant \#HST-HF2-51377 awarded by the Space Telescope Science Institute, which is operated by the Association of Universities for Research in Astronomy, Inc., for NASA, under contract NAS5-26555. JJB acknowledges support of an Australian Research Council Future Fellowship (FT180100231). JvdS is funded through a ARC Laureate Fellowship program (FL140100278). 

\section*{Data Availability} 
The data underlying this article were accessed from SAMI Data Release 2 \citep{2018MNRAS.481.2299S}, available at https://datacentral.org.au. The derived data generated in this research will be shared on reasonable request to the corresponding author.

%%%%%%%%%%%%%%%%%%%%%%%%%%%%%%%%%%%%%%%%%%%%%%%%%%


\bibliographystyle{mnras}

%%%%%%%%%%%%%%%%%%%%%%%%%%%%%%%%%%%%%%%%%%%%%%%%%%

%%%%%%%%%%%%%%%%% APPENDICES %%%%%%%%%%%%%%%%%%%%%

\appendix

%%%%%%%%%%%%%%%%%%%%%%%%%%%%%%%%%%%%%%%%%%%%%%%%%%

% Don't change these lines
\bsp	% typesetting comment
\label{lastpage}
\end{document}